\documentclass[20pt]{article}
\usepackage{setspace}
\usepackage[none]{hyphenat}
\usepackage{cite}
\usepackage{amsmath}
\usepackage{hyperref}
\usepackage[utf8]{inputenc}
\usepackage{amssymb}
\usepackage{graphicx}
\usepackage{xcolor}
\usepackage{geometry}
\usepackage{subfig}
\begin{document}
\date{}
\vspace{-3cm}
\title{A Langevin Model-Based Magneto-Optic Response SuperParamagnetic Nanoparticles Recorded with a Michelson Interferometer Setup}
\author{Syed Azer Reza and Maarij Syed}
\maketitle

\section*{Abstract}

In this paper, we provide expressions for the detected optical irradiance for magneto-optic characterization of superparamagnetic nanoparticles (SPNs) in solution imparting Faraday rotation (FR) to an optical beam passing through the sample solution. For our analysis, we assume a Langevin model for SPN samples and show the presence of odd and even harmonics for SPN samples characterized with a Michelson interferometer. This optical geometry is potentially useful in understanding nonlinear SPN FR behavior as a result of particle aggregation-based scattering effects.     

\section{Introduction}

\begin{figure}[!h]
    \centering
    \includegraphics[width=0.8\linewidth]{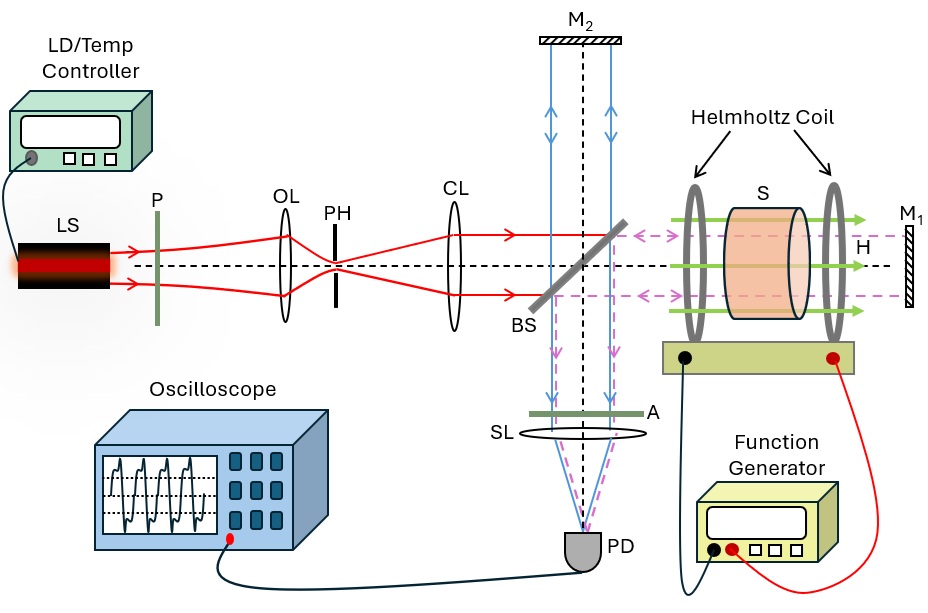}
    \caption{Interferometer setup to measure the Faraday rotation response SPN samples. Setup includes a polarizer P, an analyzer A, a pinhole PH, an objective lens OL, a collimation lens CL, sample S, a laser source LS, a spheerical lens SL, a beamsplitter BS, a photodetector PD and plane mirrors $\mathrm{M_1}$ and $\mathrm{M_2}$.}
\label{fig:SETUP}
\end{figure}

Superparamagnetic nanoparticles (SPNs) are utilized in various applications such as gene therapy \cite{uthaman2015polysaccharide}, biological sample purification \cite{weber1997specific}, drug delivery \cite{tietze2015magnetic}, contrast imaging \cite{barrow2015tailoring}, fluorescent biological labels \cite{nucci2015stem}, and microsurgical surgery. More specifically, SPNs are work agents in specific applications such as DNA extraction, contrast MRI, as well as cancer hyperthermia therapy \cite{banobre2013magnetic, kobayashi2011cancer}. Various methods have been developed to study the physical behavior of these nanoparticles \cite{miller2024multiconfiguration, roop2024magnetic, syed2024magneto}. In this study, we explore the optical response of SPNs placed under an external AC magnetic field of known frequency in an interferometric setup. More specifically, we place the SPN sample inside a quasi-uniform AC magnetic field generated by a Helmholtz coil pair. The coils are placed in one arm of a Michelson interferometer as is shown in Fig.\ref{fig:SETUP}. Light from a laser source is linearly polarized using a linear polarizer P. This linearly polarized light is then split between the two arms of the Michelson interferometer, where the secondary arm serves as a reference. Light contributions from the sample and reference arms recombine at the beamsplitter and pass through a secondary linear polarizer A (analyzer) before detection with a photodetector PD. The polarization axis of the analyzer is set to $\beta$ with respect to the polarization axis of the primary polarizer P. The PD output photocurrent is proportional to the incident optical irradiance and is measured using an oscilloscope. 

From Appendix A, the Faraday Rotation (FR) response of a SPN sample, as predicted by the Langevin model, can be described as
\begin{equation}
    \theta = VBl.
\end{equation}
More specifically, from Appendices B and C, the linear and nonlinear FR responses to an applied AC magnetic field of angular frequency $\Omega$ can be described as 
\begin{equation}
    \theta(t) = V \mu_0 H_0 (1+\chi_\mathrm{m}) l\sin\left( \Omega t \right).
    \label{eq:linearFR}
\end{equation}
and 
\begin{equation}
    \theta (t) = \mu_0 Vl \left[ (1 + \gamma M_\mathrm{s})H_0 \sin\left( \Omega t \right) - \frac{H_0^3 M_\mathrm{s}\gamma^3}{81} \sin^3\left( \Omega t \right) + \frac{H_0^5 M_\mathrm{s}\gamma^5}{5(3)^6} \sin^5\left( \Omega t \right) + \cdots \right]
    \label{eq:nonlinearFR}
\end{equation}
respectively. In Eqs.\ref{eq:linearFR} and \ref{eq:nonlinearFR}, $H_0$ is the peak amplitude of the magnetic field intensity, $\mu_0$ is the permeability of free space, $V$ is the Verdet coefficient of the sample, $\chi_\mathrm{M}$ is the magnetic susceptibility of the sample, $l$ is the sample thickness, and $M_\mathrm{s}$ is the saturation magnetization. Moreover, $\gamma = \mu/K_\mathrm{B}T$ where $\mu$ is the magnetic dipole moment of a single nanoparticle, $K$ is the Boltzmann constant and $T$ is the everage temperature of the sample.  

\section{Detected Irradiance for Polarizer-Analyzer angle setting $\beta$}
Under no E-field rotation, no light passes through the analyzer and any photodector PD, placed after the analyzer A, only records leakage 'noise' fields from the setup. But when the polarization rotates as a result of Faraday rotation, a signal is recorded at the PD which is predicted by the Malus' Law. First, let us determine the detected irradiance $\langle I_\mathrm{PD} \rangle$ for two polarizer-analyzer configurations: the $90^0$ and $45^0$ configurations. We will later use these expressions for determining the spectral decomposition of the detected irradiance for the case of linear magnetization and non-linear magnetization responses.

For a Michelson kind of measurement system, such as the one we have in the lab, the detected irradiance at the PD can be calculated as
\begin{equation}
\label{eq:int1}
    \langle I_\mathrm{PD} \rangle = \frac{E_\mathrm{1}^2}{2\eta} + \frac{E_\mathrm{2}^2}{2\eta} + \frac{1}{\eta} E_\mathrm{1}E_\mathrm{2} \cos(\Delta \psi),
\end{equation}
\begin{equation}
\label{eq:int2}
   \implies \langle I_\mathrm{PD} \rangle = I_\mathrm{1} + I_\mathrm{2} + 2\sqrt{I_\mathrm{1}I_\mathrm{2}} \cos(\Delta \psi),
\end{equation}
where $\Delta\psi$ is the phase difference due to a path length difference between the two arms of the interferometer. The respective E-field contributions $E_\mathrm{1}$ and $E_\mathrm{2}$ from the reference and sample arms arriving at the PD can be expressed as
\begin{equation}
    E_\mathrm{1} = \frac{E_0}{\sqrt{2}\sqrt{2}}\sin\left(\beta \right) = \frac{E_0}{2}\sin\left(\beta \right) ,
\end{equation}
and
\begin{equation}
\label{eq:E-field1B}
    E_\mathrm{2} = \frac{E_0}{\sqrt{2}\sqrt{2}} \sin\left( \beta + 2\theta \right) = \frac{E_0}{2} \sin\left(\beta + 2\theta \right)
\end{equation}
with $E_0$ being the initial E-field intensity from the laser source and each square root of '2' signifying one propagation through the beamsplitter. The $2\theta$ factor comes from the non-reciprocal Faraday rotation as a result of light propagating through the sample twice!

\subsection*{Special Case 1: $\beta = 45^0$ Linear Magnetization Response State}

For the measurement with a Michelson interferometer with a polarizer-analyzer angle $\beta$ set at $45^0$, we obtain
\begin{equation}
\label{eq:Field1_int}
    E_\mathrm{1} = \frac{E_0}{2\sqrt{2}},
\end{equation}
and
\begin{equation}
\label{eq:Field2_int}
E_2 = \frac{E_0}{\sqrt{2}\sqrt{2}\sqrt{2}} \left[ \cos \left( 2\theta \right) + \sin \left( 2\theta \right) \right] = \frac{E_0}{2\sqrt{2}} \left[ \cos \left( 2\theta \right) + \sin \left( 2\theta \right) \right]
\end{equation}
being the E-field contributions from the reference and sample arms respectively which make it through the analyzer 'A' and $\alpha = 2\theta$ due to double polarization rotation in a Michelson configuration. For $\Delta \psi = 0$,
\begin{equation}
    \langle I_\mathrm{PD} \rangle = \frac{E_0^2}{16\eta} + \frac{E_0^2}{16 \eta} \left[ 1 + \sin\left( 2\alpha \right) \right] + \frac{E_0^2}{8\eta} \left[ \cos \left( \alpha \right) + \sin \left( \alpha \right) \right]
\end{equation}
\begin{equation}
     \langle I_\mathrm{PD} \rangle = \frac{E_0^2}{16\eta} + \frac{E_0^2}{16 \eta} \left[ 1 + \sin\left( 4\theta \right) \right] + \frac{E_0^2}{8\eta} \left[ \cos \left( 2\theta \right) + \sin \left( 2\theta \right) \right]
\end{equation}
\begin{equation}
\label{eq:45}
     \implies \langle I_\mathrm{PD} \rangle = \frac{E_0^2}{16\eta} + \frac{E_0^2}{16 \eta} \left[ 1 + \sin\left( 4A_0 \sin\left[ \Omega t \right] \right) \right] + \frac{E_0^2}{8\eta} \left[ \cos \left( 2A_0 \sin\left[ \Omega t \right] \right) + \sin \left( 2A_0 \sin\left[ \Omega t \right] \right) \right].
\end{equation}

Applying the Jacobi-Anger expansions to Eq.\ref{eq:45} 
\begin{equation}
\label{eq:jacobi1}
\sin(z \sin \phi) \equiv 2 \sum_{ n=1 }^{\infty} J_{2n-1}(z) \sin\left[\left(2n-1\right) \phi\right],
\end{equation}
\begin{equation}
\label{eq:jacobi2}
\cos(z \sin \phi) \equiv J_0(z)+2 \sum_{n=1}^{\infty} J_{2n}(z) \cos(2n \phi),
\end{equation}
results in
\begin{align}
\begin{split}
    \langle I_\mathrm{PD} \rangle & = \frac{E_0^2}{16\eta} + \frac{E_0^2}{16\eta} + \frac{E_0^2}{8\eta}  \left[ J_1 (4A_0)\sin \left( \Omega t \right) + J_3(4A_0) \sin \left( 3\Omega t \right) + J_5 (4A_0)\sin \left( 5\Omega t  \right) + \cdots \right] \\ & + \frac{E_0^2}{8\eta} \left[ \cos\left( 2A_0\sin\left( \Omega t \right) \right) + \sin\left( 2A_0 \sin\left( \Omega t \right) \right) \right].
\end{split}
\end{align}
Expanding further
\begin{align}
\begin{split}
    \langle I_\mathrm{PD} \rangle & = \frac{E_0^2}{8\eta} + \frac{E_0^2}{8\eta}  \left[ J_1 (4A_0)\sin \left( \Omega t \right) + J_3(4A_0) \sin \left( 3\Omega t \right) + J_5 (4A_0)\sin \left( 5\Omega t  \right) + \cdots \right] \\& + \frac{E_0^2}{8\eta}  \big[ J_0 (2A_0) + 2 J_2(2A_0) \cos\left( 2\Omega t \right) + 2J_4(2A_0)\cos\left(4\Omega t \right) + \cdots + \\& 2J_1(2A_0)\sin\left( \Omega t \right) + 2J_3(2A_0) \sin\left( 3\Omega t \right) + 2J_5(2A_0) \sin\left( 5\Omega t \right) + \cdots  \big].
\end{split}
\end{align}
Rearranging the terms, we obtain
\begin{align}
    \begin{split}
        \langle I_\mathrm{PD} \rangle & =  \frac{E_0^2}{8\eta} + \frac{E_0^2}{8\eta} J_0 (2A_0) + \frac{E_0^2}{8\eta} \big[ J_1 (4A_0) + 2 J_1 (2A_0) \big] \sin\left( \Omega t \right) + \frac{E_0^2}{4\eta} J_2(2A_0) \cos\left( 2\Omega t \right) \\& + \frac{E_0^2}{8\eta} \big[ J_3(4A_0)  + 2 J_3(2A_0) \big] \sin\left( 3\Omega t \right) + \frac{E_0^2}{4\eta} J_4(2A_0)\cos\left(4\Omega t \right) + \frac{E_0^2}{8\eta} \big[ J_5(4A_0) + 2 J_5(2A_0) \big]\sin\left( 5\Omega t \right) + \cdots
    \end{split}
\end{align}

\subsection*{Special Case 2: $\beta = 45^0$ Non-linear Magnetization Response}

A non-linear magnetization response to an applied magnetic field entails a non-linear Faraday rotation to the optical polarization. As shown in Eq.\ref{eq:theta_NL}, the Faraday rotation $\theta$ is expressed as
\begin{equation}
     \theta (t) \approx \mu_0 Vl \left[ (1 + \gamma M_\mathrm{s})H_0 \sin\left( \Omega t \right) - \frac{H_0^3 M_\mathrm{s}\gamma^3}{81} \sin^3\left( \Omega t \right) \right].
\end{equation}
From \ref{eq:Field1_int} and \ref{eq:Field2_int}, for a Michelson interferometer setup with the polarizer and analyzer at $45^0$, we have
\begin{equation}
    E_\mathrm{1} = \frac{E_0}{2\sqrt{2}},
\end{equation}
and
\begin{equation}
E_2 = \frac{E_0}{\sqrt{2}\sqrt{2}\sqrt{2}} \left[ \cos \left( \alpha \right) + \sin \left( \alpha \right) \right] = \frac{E_0}{2\sqrt{2}} \left[ \cos \left( 2\theta \right) + \sin \left( 2\theta \right) \right],
\end{equation}
The nonlinear function $\theta(t)$ is 
\begin{equation}
    \theta (t) = \mu_0 Vl \left[ (1 + \gamma M_\mathrm{s})H_0 \sin\left( \Omega t \right) - \frac{H_0^3 M_\mathrm{s}\gamma^3}{81} \sin^3\left( \Omega t \right) + \frac{H_0^5 M_\mathrm{s}\gamma^5}{5(3)^6} \sin^5\left( \Omega t \right) + \cdots \right].
\end{equation}
The number of terms that we consider in this non-linear expansion depends on the values of $M_\mathrm{s}$ and $\gamma = \mu\mu_0/K_\mathrm{B}T$, as well as the value of the amplitude $H_0$ of the applied magnetic field. If $H_0 M_\mathrm{s} \mu / K_\mathrm{B} T << 1$, then we can possibly consider the third order order term only to approximate the non-linear magnetization response. In this case, we can state that 
\begin{equation}
     \theta (t) \approx \mu_0 Vl \left[ (1 + \gamma M_\mathrm{s})H_0 \sin\left( \Omega t \right) - \frac{H_0^3 M_\mathrm{s}\gamma^3}{81} \sin^3\left( \Omega t \right) \right].
\end{equation}
Using this $\theta(t)$ into Eq.\ref{eq:Irrad_int_NL}, we obtain the time-averaged irradiance recorded by a photodetector. With
\begin{equation}
    \kappa_1 = 2\mu_0 Vl H_0 (1+\gamma M_\mathrm{s}),
\end{equation}
and 
\begin{equation}
    \kappa_3 = -\frac{2\mu_0 VlH_0^3 M_\mathrm{s}\gamma^3}{81},
\end{equation}
signifying the amplitudes of the $\sin(\Omega t)$ and $\sin(3\Omega t)$ terms, 
\begin{equation}
\label{eq:theta_Int}
     \theta (t) \approx \frac{\kappa_1}{2} \sin\left( \Omega t \right) + \frac{\kappa_3}{2} \sin^3\left( \Omega t \right) .
\end{equation}
The time-averaged irradiance 
\begin{equation}
\label{eq:Irrad_int_NL}
     \langle I_\mathrm{PD} \rangle = \frac{|E_\mathrm{PD}|^2}{2\eta} = \frac{|E_1 + E_2|^2}{2\eta} = \frac{E_0^2}{16\eta} + \frac{E_0^2}{16 \eta} \left[ 1 + \sin\left( 4\theta \right) \right] + \frac{E_0^2}{8\eta} \left[ \cos \left( 2\theta \right) + \sin \left( 2\theta \right) \right].
\end{equation}
Substituting for $\theta$ from Eq.\ref{eq:theta_Int}, we obtain 
\begin{align}
    \begin{split}
        \langle I_\mathrm{PD} \rangle & = \frac{E_0^2}{16\eta} + \frac{E_0^2}{16 \eta} \left\{ 1 + \sin\left[ 2\kappa_1\sin\left(\Omega t \right) + 2\kappa_3 \sin^3(\Omega t) \right] \right\} + \\& \frac{E_0^2}{8\eta} \left\{ \cos \left[ \kappa_1 \sin(\Omega t) + \kappa_3 \sin^3(\Omega t)  \right] + \sin \left[ \kappa_1 \sin(\Omega t) + \kappa_3 \sin^3(\Omega t) \right] \right\}.
    \end{split}
\end{align}
Applying the Triple-Angle Identity, we get
\begin{equation}
    \sin^3\left( \Omega t \right) = \frac{3\sin\left( \Omega t \right) - \sin\left(3\Omega t \right)}{4},
\end{equation}
we arrive at 
\begin{align}
\label{eq:IPD_NL_Int}
    \begin{split}
        \langle I_\mathrm{PD} \rangle  = & ~ \frac{E_0^2}{16\eta} + \frac{E_0^2}{16 \eta} \left\{ 1 + \sin\left[ 2\kappa_1 \sin\left( \Omega t \right) + \frac{3}{2} \kappa_3 \sin\left( \Omega t \right) - \frac{1}{2} \kappa_3 \sin\left(3\Omega t \right)  \right] \right\} + \\& + \frac{E_0^2}{8\eta} \left\{ \cos \left[ \kappa_1 \sin\left( \Omega t \right) + \frac{3}{4} \kappa_3 \sin\left( \Omega t \right) - \frac{1}{4} \kappa_3 \sin\left(3\Omega t \right)   \right] \right\} + \\& + \frac{E_0^2}{8\eta} \left\{ \sin \left[ \kappa_1 \sin\left( \Omega t \right) + \frac{3}{4} \kappa_3 \sin\left( \Omega t \right) - \frac{1}{4} \kappa_3 \sin\left(3\Omega t \right) \right] \right\} .
    \end{split}
\end{align}
With $A_1 = \kappa_1 + 3\kappa_3/4 $ and $A_3 = - \kappa_3/4$, we express Eq.\ref{eq:IPD_NL_Int} as
\begin{align}
\label{eq:IPD_NL_Int2}
    \begin{split}
        \langle I_\mathrm{PD} \rangle  = & ~ \frac{E_0^2}{16\eta} + \frac{E_0^2}{16 \eta} \left\{ 1 + \sin\left[ 2A_1 \sin\left( \Omega t \right) + 2A_3 \sin\left(3\Omega t \right)  \right] \right\} + \\& + \frac{E_0^2}{8\eta} \left\{ \cos \left[ A_1 \sin\left( \Omega t \right) + A_3 \sin\left(3\Omega t \right) \right] \right\} + \\& + \frac{E_0^2}{8\eta} \left\{ \sin \left[ A_1 \sin\left( \Omega t \right) + A_3 \sin\left(3\Omega t \right) \right] \right\} .
    \end{split}
\end{align}
This leads to
\begin{align}
    \begin{split}
        \langle I_\mathrm{PD} \rangle  = & ~ \frac{E_0^2}{16\eta} + \frac{E_0^2}{16 \eta} + \frac{E_0^2}{16 \eta} \left\{\sin\left[ 2A_1\sin(\Omega t)\right] \cos\left[ 2A_3\sin(3\Omega t) \right] + \cos\left[ 2A_1\sin(\Omega t)\right] \sin\left[ 2A_3\sin(3\Omega t) \right] \right\} \\& + \frac{E_0^2}{8\eta} \left\{ \cos\left[ A_1\sin(\Omega t)\right] \cos\left[ A_3\sin(3\Omega t) \right] - \sin\left[ A_1\sin(\Omega t)\right] \sin\left[ A_3\sin(3\Omega t) \right] \right\} \\& + \frac{E_0^2}{8\eta}\left\{ \sin\left[ A_1\sin(\Omega t)\right] \cos\left[ A_3\sin(3\Omega t) \right] + \cos\left[ A_1\sin(\Omega t)\right] \sin\left[ A_3\sin(3\Omega t) \right] \right\}
    \end{split}
\end{align}
Using the Jacobi-Anger Expansion formulas
\begin{equation}
\sin(z \sin \phi) \equiv 2 \sum_{ n=1 }^{\infty} J_{2n-1}(z) \sin\left[\left(2n-1\right) \phi\right],
\end{equation}
\begin{equation}
\cos(z \sin \phi) \equiv J_0(z)+2 \sum_{n=1}^{\infty} J_{2n}(z) \cos(2n \phi),
\end{equation}
\begin{align}
    \begin{split}
        \langle I_\mathrm{PD} \rangle  = & ~ \frac{E_0^2}{8\eta} +  \\& 
        + \frac{E_0^2}{4 \eta} \left\{\left[J_1(2A_1)\sin\left(\Omega t \right) + J_3(2A_1)\sin\left(3\Omega t \right) \right]\left[ \frac{J_0(2A_3)}{2} + J_2(2A_3)\cos(6\Omega t) + J_4(2A_3)\cos(12\Omega t) \right] \right\} 
        \\& 
        + \frac{E_0^2}{4 \eta} \left\{\left[J_1(2A_3)\sin\left(3\Omega t \right) + J_3(2A_3)\sin\left(9\Omega t \right) \right] \left[\frac{J_0(2A_1)}{2} + J_2(2A_1)\cos(2\Omega t) + J_4(2A_1)\cos(4\Omega t) \right] \right\} 
        \\&  
        + \frac{E_0^2}{2\eta} \left\{ \frac{J_0(A_1)}{2} + J_2(A_1) \cos(2\Omega t) + J_4(A_1) \cos(4\Omega t) + J_6(A_1) \cos(6\Omega t) + \cdots \right\} \cdot 
        \\& 
        \cdot \left\{ \frac{J_0(A_3)}{2} + J_2(A_3) \cos(6\Omega t) + J_4(A_3) \cos(12\Omega t) + \cdots \right\} 
        \\&  
        - \frac{E_0^2}{2\eta} \left\{  J_1(A_1) \sin(\Omega t) + J_3(A_1) \sin(3\Omega t) + J_5(A_1) \sin(5\Omega t) + \cdots \right\} \cdot 
        \\& 
        \cdot \left\{  J_1(A_3) \sin(3\Omega t) + J_3(A_3) \sin(9\Omega t)  + \cdots \right\}
        \\&
        + \frac{E_0^2}{2 \eta} \left\{\left[J_1(A_1)\sin\left(\Omega t \right) + J_3(A_1)\sin\left(3\Omega t \right) \right]\left[ \frac{J_0(A_3)}{2} + J_2(A_3)\cos(6\Omega t) + J_4(A_3)\cos(12\Omega t) \right] \right\} 
        \\& 
        + \frac{E_0^2}{2 \eta} \left\{\left[J_1(A_3)\sin\left(3\Omega t \right) + J_3(A_3)\sin\left(9\Omega t \right) \right] \left[\frac{J_0(A_1)}{2} + J_2(A_1)\cos(2\Omega t) + J_4(A_1)\cos(4\Omega t) \right] \right\} 
    \end{split}
    \label{eq:full_exp_NL}
\end{align}
As typical values of $\theta $ are assumed low (even for SPM nanoparticle samples which impart a higher Faraday rotation), we assume that for all practical purposes, we can ignore the Bessel function terms $J_n$ for $n \geq 3$. We also only evaluate the first four harmonics (up to $4\Omega$). These simplifications allow us to have a much simplified approximation to our expansion in Eq.\ref{eq:full_exp_NL}. Keeping only the terms which fulfill these conditions, we obtain
\begin{align}
    \begin{split}
        \langle I_\mathrm{PD} \rangle  = & ~ \frac{E_0^2}{8\eta} \\& 
        + \frac{E_0^2}{4 \eta} \left\{ \frac{J_0(2A_3)J_1(2A_1)}{2} \sin(\Omega t) + \frac{J_0(2A_3)J_3(2A_1)}{2} \sin(3\Omega t) + \frac{J_0(2A_1)J_1(2A_3)}{2} \sin(3\Omega t)  \right\} 
        \\&
        + \frac{E_0^2}{4 \eta} \left\{ \frac{J_1(2A_3)J_2(2A_1)}{2} \sin(\Omega t) - \frac{J_3(2A_1)J_2(2A_3)}{2}\sin(3\Omega t) + \frac{J_1(2A_3)J_2(2A_1)}{2} \sin(5\Omega t)  \right\}
        \\&
        + \frac{E_0^2}{2 \eta} \left\{ \frac{J_0(A_1)J_0(A_3)}{4} + \frac{J_0(A_3)J_2(A_1)}{2}\cos(2\Omega t) - \frac{J_1(A_1) J_1(A_3)}{2} \cos(2\Omega t) + \frac{J_1(A_1) J_1(A_3)}{2} \cos(4\Omega t) \right\} 
        \\&
        + \frac{E_0^2}{2 \eta} \left\{ \frac{J_2(A_1)J_2(A_3)}{2}\cos(4\Omega t) - J_3(A_1)J_1(A_3) + J_4(A_1)J_2(A_3)\cos(2\Omega t) \right\}
        \\&
        + \frac{E_0^2}{2 \eta} \left\{ \frac{J_0(A_3)J_1(A_1)}{2} \sin(\Omega t) + \frac{J_0(A_3)J_3(A_1)}{2} \sin(3\Omega t) + \frac{J_0(A_1)J_1(A_3)}{2} \sin(3\Omega t)   \right\} 
        \\&
        + \frac{E_0^2}{2 \eta} \left\{ \frac{J_1(A_3)J_2(A_1)}{2} \sin(\Omega t) - \frac{J_3(A_1)J_2(A_3)}{2}\sin(3\Omega t) + \frac{J_1(A_3)J_2(A_1)}{2} \sin(5\Omega t) \right\} 
    \end{split}
\end{align}
Rearranging these terms leads to
\begin{align}
    \begin{split}
        & \langle I_\mathrm{PD} \rangle  = \frac{E_0^2}{8\eta} + \frac{J_0(A_1)J_0(A_3) E_0^2}{8\eta} - \frac{J_3(A_1)J_1(A_3)}{2\eta}
        \\&
        + \frac{E_0^2}{\eta} \left\{ \frac{J_0(2A_3)J_1(2A_1) }{8} + \frac{J_1(2A_3)J_2(2A_1) }{8} + \frac{J_0(A_3)J_1(A_1) }{4} + \frac{J_1(A_3)J_2(A_1) }{4} + \cdots \right\} \sin(\Omega t)
        \\&
        + \frac{E_0^2}{\eta} \left\{ \frac{J_0(A_3)J_2(A_1)}{4} - \frac{J_1(A_1) J_1(A_3)}{4} + \frac{J_4(A_1)J_2(A_3) }{4} +\cdots \right\}\cos(2\Omega t)
        \\&
        + \frac{E_0^2}{\eta} \left\{  \frac{J_0(2A_3)J_3(2A_1)}{8} + \frac{J_0(2A_1)J_1(2A_3)}{8}  + \frac{J_0(A_3)J_3(A_1)}{4}  + \frac{J_0(A_1)J_1(A_3)}{4} +\cdots  \right\} \sin(3\Omega t)
        \\&
        - \frac{E_0^2}{ \eta} \left\{ \frac{J_3(2A_1)J_2(2A_3)}{8} + \frac{J_3(A_1)J_2(A_3)}{4} + \cdots \right\}\sin(3\Omega t)
        \\&
        + \frac{E_0^2}{\eta} \left\{ \frac{J_1(A_1) J_1(A_3)}{4} + \frac{J_2(A_1)J_2(A_3)}{4} + \cdots \right\} \cos(4\Omega t)
    \end{split}
\end{align}

After this separation of harmonic contributions, we replace $A_1$ and $A_3$ in terms of $\theta$ to get a sense of the contributions from each of the terms.

\newpage

\subsection*{Appendix A: Faraday Rotation of Polarization}

\begin{figure}[!h]
    \centering
    \includegraphics[width=0.8\linewidth]{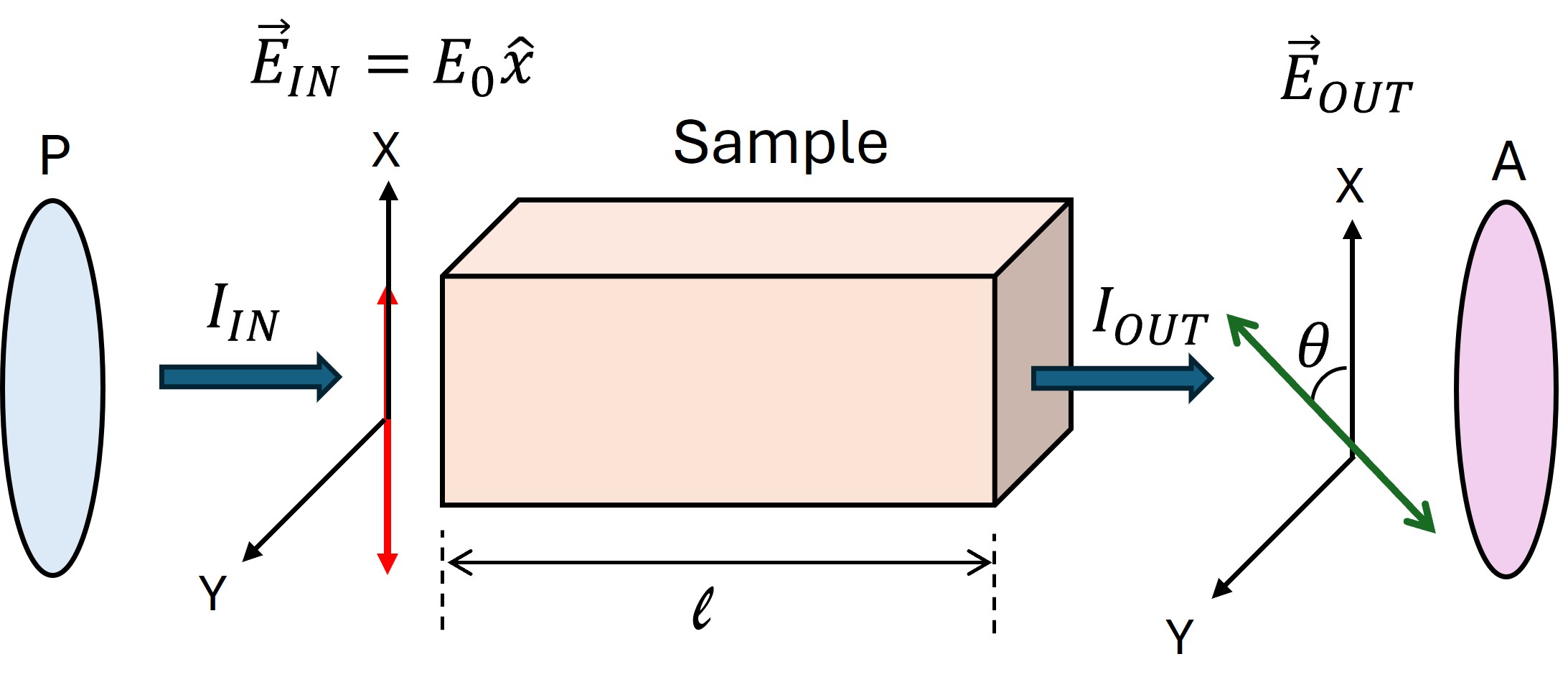}
    \caption{Polarization-based measurements with an Faraday Faraday rotation $\theta$}
\label{fig:FR3}
\end{figure}

In this appendix, we present basic calculations that allow us to analyze the performance of a single-pass optical setup that is used for measuring Faraday rotation in magneto-optic samples. A typical setup is shown in Fig.\ref{fig:FR3} where an initial polarizer P linearly polarizes light, which then passes through the sample which imparts an angular rotation $\theta$ to the incident polarization set by P. 
The light exiting the sample then passes through a second polarizer A which we refer to as an analyzer. The analyzer is placed after the sample so that its axis of polarization is at $\beta$ with respect to 'P'. The changes in the recorded irradiance by a photodetector can then be used to determine the polarization rotation $\theta$ due to the magneto-optic effect known as Faraday rotation. Without the sample present, the time-averaged recorded irradiance by the photodetector is described by Malus Law as
\begin{equation}
    \langle I_\mathrm{PD} \rangle \propto \cos^2\beta.
\end{equation}
The initial polarizer P linearly polarizes light along a single axis. The linearly polarized optical E-field of a plane wave, propagating along the $+z$ axis can be described as a phasor
\begin{equation}
    \Vec{E} = \Vec{E_0}e^{j(\omega t - kz)},
\end{equation}
where $\omega$ is the optical angular frequency, $k = \omega/c$ is the wave number and $E_0$ is the peak oscillation amplitude of the E-field. For linear polarization in the x-y plane (say along the x-axis), we can state that $\Vec{E_0} = E_0 \hat{i}$.
\begin{figure}
    \centering
    \includegraphics[width=0.5\linewidth]{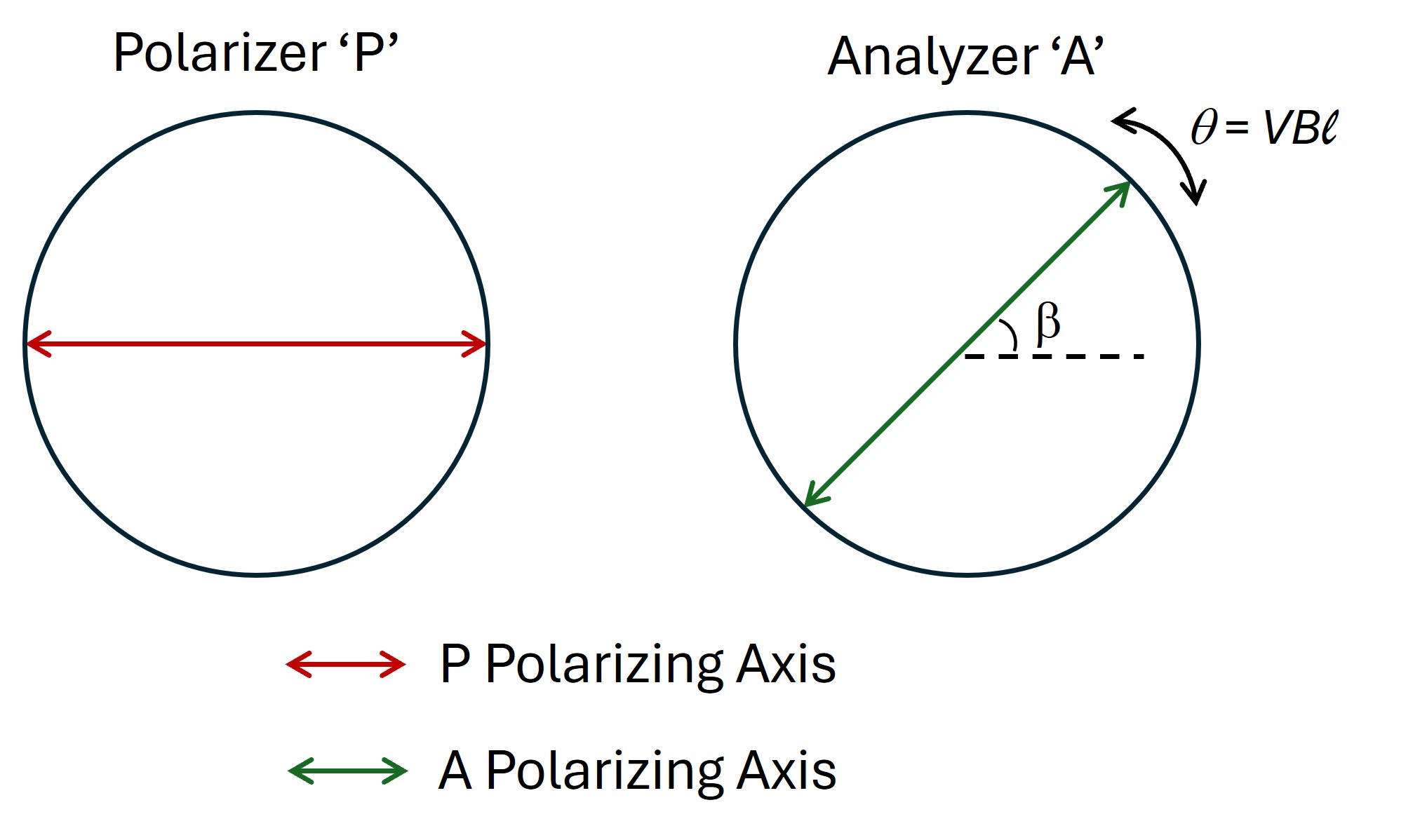}
    \caption{A general polarizer-analyzer setting with a mutual polarization angle $\beta$}
\label{Fig:FR2A}
\end{figure}
For nonzero Faraday rotation $\theta$, the optical Electric field (as well as the optical magnetic field) of light rotates in the presence of an external longitudinal applied magnetic field which produced a magnetic flux density magnitude of $|\Vec{B}| = B$. The resulting Faraday rotation is given by
\begin{equation}
    \theta = VBl,
\end{equation}
where $V$ is the Verdet constant of the sample and '$l$' is the longitudinal dimension of the sample through which light propagates. This is shown in Figure.\ref{Fig:FR2A} where the polarization of the optical field rotates as a function of the applied external magnetic field inside which the sample is placed. For a polarizer-analyzer combination with linear polarization axes angled at $\beta$ with respect to each other, the E-field $E_\mathrm{PD}$ transmitted through the analyzer is recorded by a photodetector. Here
\begin{equation}
    E_\mathrm{PD} = E_0 \cos\left( \beta + \theta \right)
\end{equation}
as shown in Fig.\ref{Fig:FR2A}. The resulting time-averaged irradiance, recorded by the PD, is
\begin{equation}
\label{eq:Irr_efield}
   \langle I_\mathrm{PD} \rangle = \frac{|E_\mathrm{PD}|^2}{2\eta}
\end{equation}
where $\eta$ is the impedance of air.

\subsection*{Appendix B: Linear Faraday Rotation Response to Applied Sinusoidal Magnetic Field}

The most common polarization rotation imparted to a linearly polarized propagating optical wave through a sample is due to a linear magnetization response of the sample to the applied magnetic field intensity $H$. In the context of SPM nanoparticles, a low particle concentration should also lead to a response that somewhat mimics a linear magnetization response. This implies that the resulting magnetic flux density $B$ produced within the sample is linearly proportional to the applied magnetic field $H$. Therefore, $B$ within the magnetically linear sample is expressed as
\begin{equation}
    B = \mu_0 H + \mu_0 M = \mu_0 H + \mu_0 \chi_\mathrm{m} H = \mu_0 H, (1+\chi_\mathrm{m}),
\end{equation}
where $M = \chi_\mathrm{m} H$ is the samples' linear magnetization and $\mu_r = 1+\chi_\mathrm{m}$ is the magnetic permeability of the sample medium in relation to the permeability of vacuum $\mu_0$. In the case where the applied magnetic field $H (t)$ to the sample is a time-varying signal, such as a sinusoidal signal of angular frequency $\Omega$ of zero mean
\begin{equation}
    H(t) = H_0\sin \left( \Omega t \right).
\end{equation}
 The corresponding flux density is
\begin{equation}
    B(t) =  B_0 \sin \left( \Omega t \right),
\end{equation}
with $B_0 = \mu_0 H_0 (1+\chi_\mathrm{m})$. This results in a time-modulated Faraday rotation which is expressed as
\begin{equation}
    \theta(t) = VB_0 l\sin\left( \Omega t \right).
\end{equation}

\subsection*{Appendix C: Nonlinear Faraday Rotation Response to Applied Sinusoidal Magnetic Field}

Magnetization for mainly paramagnetic or diamagnetic materials is mostly linear in relation to the applied magnetic field intensity $H$. In such cases, the relation between the applied magnetic field $H$ and the resulting magnetic flux density $B$ is described simply as a function of the material susceptibility $\chi_\mathrm{M}$ (for magnetically isotropic materials, this is simply a scalar number) as:
\begin{equation}
    B = \mu_0 H + \mu_0 M,
\end{equation}
where $M = \chi_\mathrm{M} H$. Therefore,
\begin{equation}
    B = \mu_0 H\left( 1+\chi_\mathrm{M} \right),
\end{equation}
and the corresponding Faraday rotation for magneto-optically active materials is described as
\begin{equation}
    \theta = VL\mu_0 H\left( 1 + \chi_\mathrm{M} \right).
\end{equation}
In contrast, the relation between the magnetization and the applied magnetic field is nonlinear for superparamagnetic (SPM) materials. It is described as
\begin{equation}
    M = M_\mathrm{s} L\left( H\frac{\mu\mu_0}{K_\mathrm{B}T}\right), 
\end{equation}
where $L(x)$ is the Langevin function of argument $x$ described by
\begin{equation}
    L(x) = \frac{1}{\tanh\left( x \right)} - \frac{1}{x},
\end{equation}
and $M_\mathrm{s}$ is the saturation magnetization of the material which is expressed as 
\begin{equation}
    M_\mathrm{s} = N\mu,
\end{equation} 
where $N$ is the nanoparticle concentration per unit volume and $\mu$ is the magnetic moment of a single nanoparticle. Consequently, the magnetization can be simply stated as
\begin{equation}
    M = M_\mathrm{s} L(\gamma H) = M_\mathrm{s} \left[ \frac{1}{\tanh\left( \gamma H \right)} - \frac{1}{\gamma H} \right]
\end{equation}
with $\gamma = \mu_0\mu/K_\mathrm{B} T$. Typically a Laurent expansion can be used to expand the Langevin function but the resulting series has a non-trivial mathematical form. For simplicity, we use a much simpler approximation of the Langevin function, where 
\begin{equation}
    L(\gamma H) \approx \tanh \left( \frac{1}{3} \gamma H \right),
\end{equation}
Hence the resulting magnetization can be approximation can be described as
\begin{equation}
    M = M_\mathrm{s} \tanh \left( \frac{1}{3} \gamma H \right) = M_\mathrm{s} \tanh \left( \frac{\mu\mu_0}{3K_\mathrm{B}T} H \right) 
\end{equation}
The corresponding magnetic flux density $B$ is expressed as
\begin{equation}
    B = \mu_0 H + \mu_0 M_\mathrm{s} \tanh \left( \frac{\mu\mu_0}{3K_\mathrm{B}T} H \right)
\end{equation}
We now express the hyperbolic tangent function into its MacLaurin series expansion as
\begin{equation}
    \tanh \left( x \right) = x - \frac{x^3}{3} + \frac{2x^5}{15} + \cdots 
\end{equation}
This results in 
\begin{equation}
    B = \mu_0 H + \mu_0 M_\mathrm{s} \left[ \frac{\gamma H}{3} - \frac{(\gamma H)^3}{3^4} + \frac{(\gamma H)^5}{5(3)^6} + \cdots \right]
\end{equation}
When applying a sinusoidal H field $H = H_0\sin\left( \Omega t \right)$, we obtain a Faraday rotation of
\begin{equation}
    \theta = VBl = Vl \left\{ \mu_0 H + \mu_0 M_\mathrm{s} \left[ \frac{\gamma H}{3} - \frac{\gamma^3 }{81} H^3 + \frac{\gamma^5}{5(3)^6} H^5 + \cdots \right] \right\}
\end{equation}
\begin{equation}
    \theta = \mu_0 Vl \left[ (1+\gamma M_\mathrm{s}) H - \frac{M_\mathrm{s}\gamma^3}{81} H^3 + \frac{M_\mathrm{s}\gamma^5}{5(3)^6} H^5 + \cdots \right].
\end{equation}
Therefore,
\begin{equation}
    \theta (t) = \mu_0 Vl \left[ (1 + \gamma M_\mathrm{s})H_0 \sin\left( \Omega t \right) - \frac{H_0^3 M_\mathrm{s}\gamma^3}{81} \sin^3\left( \Omega t \right) + \frac{H_0^5 M_\mathrm{s}\gamma^5}{5(3)^6} \sin^5\left( \Omega t \right) + \cdots \right]
\end{equation}
The number of terms that we consider in this non-linear expansion depends on the values of $M_\mathrm{s}$ and $\gamma = \mu/K_\mathrm{B}T$, as well as the value of the amplitude $H_0$ of the applied magnetic field. If $H_0 M_\mathrm{s} \mu / K_\mathrm{B} T << 1$, then we can possibly consider the third-order term only to approximate the non-linear magnetization response. In this case, we can state that 
\begin{equation}
\label{eq:theta_NL}
     \theta (t) \approx \mu_0 Vl \left[ (1 + \gamma M_\mathrm{s})H_0 \sin\left( \Omega t \right) - \frac{H_0^3 M_\mathrm{s}\gamma^3}{81} \sin^3\left( \Omega t \right) \right].
\end{equation}

\bibliographystyle{IEEEtran}
\bibliography{poynting.bib}

\end{document}